\documentclass{article}

\usepackage{microtype}
\usepackage{graphicx}
\usepackage{subcaption}
\usepackage{booktabs} 

\usepackage{hyperref}

\usepackage[preprint]{icml2026}

\usepackage{amsmath}
\usepackage{amssymb}
\usepackage{mathtools}
\usepackage{amsthm}
\usepackage{ulem}
\usepackage{url}
\usepackage[f]{esvect}
\usepackage{multirow}

\usepackage{bigints}
\usepackage{amsfonts,amsmath,amssymb,amsthm, bbm, bm, mathtools}

\DeclareMathOperator*{\argmin}{arg\,min}

\usepackage{tikz}
\usetikzlibrary{automata, arrows.meta, positioning, fit, shapes}

\usepackage[capitalize,noabbrev]{cleveref}

\theoremstyle{plain}

\theoremstyle{definition}

\theoremstyle{remark}

\usepackage[textsize=tiny]{todonotes}

\icmltitlerunning{Change Point Detection for Cell Populations Measured via Flow Cytometry}

\begin{document}

\twocolumn[
  \icmltitle{Change Point Detection for Cell Populations Measured via Flow Cytometry}



  \icmlsetsymbol{equal}{*}

  \begin{icmlauthorlist}
    \icmlauthor{Yik Lun Kei}{equal,UCSC}
    \icmlauthor{Qi Wang}{equal,UCSC}
    \icmlauthor{Paul A. Parker}{UCSC}
    \icmlauthor{Francois Ribalet}{UW}
    \icmlauthor{Sangwon Hyun}{UCSC}
  \end{icmlauthorlist}

  \icmlaffiliation{UCSC}{Department of Statistics, University of California, Santa Cruz}
  \icmlaffiliation{UW}{School of Oceanography, University of Washington}

  \icmlcorrespondingauthor{Sangwon Hyun}{shyun2@ucsc.edu}

  \icmlkeywords{Change Point Detection, Flow Cytometry, Group-fused LASSO, Latent Space Models, Replicated Single-cell Data}

  \vskip 0.3in
]



\printAffiliationsAndNotice{}  

\begin{abstract}
The ocean is filled with phytoplankton that contribute as much photosynthesis as all land plants combined, making them vital to the carbon cycle and climate system. Recent advances in flow cytometry allow oceanographers to measure the optical traits of individual cells along research cruise tracks, generating single-cell resolution microbial data. In marine microbial ecology, detecting locations of abrupt changes in the environmental response of cytometric plankton distributions is an important task. This manuscript proposes a latent space Gaussian mixture-of-experts model, facilitating change point detection in replicated and clustered phytoplankton observations. Change points are identified through shifts in prior means of low-dimensional representations, with piecewise-constant structure enforced by a group-fused LASSO penalty. The optimization problem is then solved via Alternating Direction Method of Multipliers. Applied to flow cytometry data, the proposed method identifies a scientifically important change point that aligns with a transition zone between two marine provinces.
\end{abstract}

\section{Introduction}

Phytoplankton are microscopic photosynthetic organisms that establish the foundation of marine food webs and contribute substantially to global oxygen production \citep{field1998primary}. Analyzing how phytoplankton communities vary is central to understanding ocean biochemical processes, ecosystem dynamics, and long term climate variability \citep{Ribalet2025}. Recent advances in flow cytometry have enabled continuous collection of phytoplankton single-cell measurements along a cruise track \citep{sosik2007automated,olson2017imaging}. At each hourly time point, a massive number of cells are measured, each characterized by three optical features: red fluorescence, orange fluorescence, and cell diameter. These measurements form high-frequency time series of high-dimensional observations that capture variations in phytoplankton community structure across environmental gradients \citep{hyun2023modeling}. However, the high dimensionality, natural clustering of cell populations, and extensive replication in these flow cytometry data pose significant challenges for statistical analysis.

One fundamental task in analyzing flow cytometry time series is to identify locations where the distribution of phytoplankton species and their relationship with environmental conditions undergo abrupt changes. These changes may correspond to transitions between water masses, shifts in nutrient availability, and other environmental forcing. In practice, distributional changes in flow cytometry data are often heterogeneous and intertwined. Accurately pinpointing these change points therefore provides a data-driven way to delineate environmental transitions and to characterize where ocean conditions shift.

In recent decades, myriad change point detection methods have been proposed for univariate and multivariate time series without replications and covariates. \cite{bai1997estimating} formulated the problem as detecting structural breaks with linear models, identifying change points through least-squares estimation of piecewise-constant means. \cite{killick2012optimal} developed the pruned exact linear time algorithm, which achieves multiple change point detection using a penalized likelihood formulation with linear computational complexity. \cite{matteson2014nonparametric} proposed a non-parametric divisive procedure, which detects distributional changes by recursively partitioning the time series using multivariate distance measures. Similarly, \cite{chen2015graph}, \cite{chu2019asymptotic}, and \cite{song2022asymptotic} developed non-parametric methods that characterize change points through discrepancies between empirical distributions across time. Moreover, \cite{garreau2018consistent} and \cite{song2022new} employed kernel-based methods to extract features in high dimensions that delineate the distributional differences.

However, many existing change point detection methods are not directly applicable to flow cytometry data due to several structural challenges. First, the natural clustering of cells into distinct populations (i.e., species) induces a complex and multimodal mixture distribution observed at a time. Second, the measurements are high-dimensional and replicated at each time point, whereas most existing methods assume a single observation per time point. Third, environmental covariates, such as salinity and light intensity, can influence both the mixture composition and the characteristics of individual populations, making it difficult to specify for a detection method which aspects drive the changes. Consequently, there is a high demand for a change point detection method that can handle replicated and clustered distributions while incorporating covariate effects.

To address these challenges, we propose a decoder-only latent space mixture model to facilitate change point detection in high-frequency replicated cell measurement data. Specifically, our framework models the conditional distribution of single-cell observations given environmental covariates, using a multivariate Gaussian mixture-of-experts decoder. The mixture weights, component means, and variances are parameterized by neural networks with observed covariates and a low-dimensional representation as input that shared by all cells observed at each time point. The low-dimensional representations are managed by Gaussian priors, whose mean parameters are learned from the flow cytometry data. Subsequently, change points are identified through shifts in the time-varying prior means, with a piecewise constant structure enforced by a group-fused LASSO penalty. This formulation enables joint learning of the decoder and prior means, while explicitly accounting for the covariate effects. The resulting optimization problem is solved via an Alternating Direction Method of Multipliers (ADMM) procedure. To the best of our knowledge, this is the first method that can directly perform change point detection on multivariate, replicated, clustered, and covariate-dependent data at the cell-population level.

The balance of this manuscript is organized as follows. Section \ref{sec_model} introduces our latent space mixture model and formulates the change point detection problem. Section \ref{sec_learning} presents the ADMM learning algorithm and Langevin dynamics. Section \ref{sec_guidance} describes procedures for change point localization and model selection. Section \ref{simulation_study} evaluates our method on synthetic data with real covariates. Section \ref{Application} demonstrates an application to real flow cytometry data from an oceanographic cruise. Section \ref{sec_discussion} concludes with a discussion of limitations and future directions.

\section{Latent Space Model}
\label{sec_model}

\subsection{Model Specification}

We consider flow cytometry data collected during an oceanographic cruise, where environmental conditions and single-cell measurements are recorded sequentially over time. At each time point $t$, we collect $\{\bm{y}_b^t\}_{b=1}^{B} \subset \mathbb{R}^3$, which correspond to three optical measurements (red fluorescence, orange fluorescence, and cell diameter) from $B$ individual cells. Let $\bm{x}^t \in \mathbb{R}^p$ denote the observed covariates at time $t$, such as temperature, sea surface salinity, and nitrite level. All cells, regardless of their species, share the same covariates $\bm{x}^t$, reflecting common environmental conditions under which the cells are sampled from the ocean. Moreover, we assume that these cells can be clustered into $K$ categories corresponding to different species.

For $\bm{y}^t \in \mathbb{R}^3$ at a specific time $t$, we consider a latent space  Gaussian mixture-of-experts decoder:
\begin{equation}
\begin{split}
\label{GoM}
\bm{y}^t \sim P(&\bm{y}^t|\bm{x}^t,\bm{z}^t) = \sum_{k=1}^K \bm{\pi}_k (\bm{x}^t,\bm{z}^t) \times\\
& \mathcal{N}^{(k)}\big(\bm{y}^t; \bm{\beta}^{(k)}(\bm{x}^t,\bm{z}^t), \bm{\Sigma}^{(k)}(\bm{x}^t,\bm{z}^t) \big)
\end{split}
\end{equation}
where $K \in \mathbb{N}$ is the number of clusters and $K$ is assumed to be known. Since the goal is to detect change points instead of clustering individual cells, a reasonable $K$ based on scientific input can be supplied to achieve the detection task. In flow cytometry, cell populations are often clustered, so a Mixture of Gaussian distributions serves as a natural and powerful model for replicated cells of various species. Here, $\bm{z}^t \in \mathbb{R}^d$ denotes the low-dimensional representation at time $t$, governing the unobserved relationship between ocean conditions and cell populations that induces changes. Further details are discussed later in this section.

Three modeling components arise from the proposed framework, and neural networks are used to promote flexibility. In particular, we parameterize the mixing coefficients of the decoder in (\ref{GoM}) as
$$\bm{\pi}(\bm{x}^t,\bm{z}^t) = \text{Softmax}\big(\text{MLP}_{\bm{\phi}_1}(\bm{x}^t,\bm{z}^t)\big) \in [0,1]^K,$$
ensuring that $\sum_{k=1}^K \bm{\pi}_k(\bm{x}^t,\bm{z}^t) = 1$. Moreover, the means of the Gaussian components are modeled as
$$\bm{\beta}(\bm{x}^t,\bm{z}^t) = \text{MLP}_{\bm{\phi}_2}(\bm{x}^t,\bm{z}^t)\ \in \mathbb{R}^{K \times 3}$$
where the $k$-th row corresponds to $\bm{\beta}^{(k)}(\bm{x}^t,\bm{z}^t) \in \mathbb{R}^3$. Lastly, the diagonal covariance matrix is
$$\text{diag}\big(\bm{\Sigma}^{(k)}(\bm{x}^t,\bm{z}^t)\big) = \text{Softplus}\big( \text{MLP}_{\bm{\phi}_3}(\bm{x}^t,\bm{z}^t)\big) \in \mathbb{R}_+^d$$
for each cluster $k \in [K]$. Throughout, $\bm{\phi} = \{\bm{\phi}_1, \bm{\phi}_2, \bm{\phi}_3\}$ denotes the collection of the neural network parameters from the three branches.

Note that the distribution of interest is the conditional density $P(\bm{y}^t | \bm{x}^t)$, and we parameterize it as
\begin{align*}
P(\bm{y}^t | \bm{x}^t) & = \int P(\bm{y}^t, \bm{z}^t | \bm{x}^t) d\bm{z}^t \\
& = \int P(\bm{y}^t| \bm{z}^t ,\bm{x}^t) P(\bm{z}^t) d\bm{z}^t.
\end{align*}
In this work, we let $P(\bm{z}^t | \bm{x}^t) \coloneqq P(\bm{z}^t)$ be data agnostic to the covariates \citep{bowman2016generating, lee2017desire, bhattacharyya2018accurate}, enforcing that the latent representation $\bm{z}^t \in \mathbb{R}^d$ abstracts ocean information from optical features $\{\bm{y}_b^t\}_{b=1}^{B} \subset \mathbb{R}^3$, conditional on ocean covariates $\bm{x}^t \in \mathbb{R}^p$ at time $t$. 

Subsequently, we impose a learnable Gaussian prior to the latent variable as
$$\bm{z}^t \sim P(\bm{z}^t) = \mathcal{N}(\bm{z}^t; \bm{\mu}^t, \bm{I}_d)$$ where $\bm{\mu}^t \in \mathbb{R}^d$ is the mean vector and $\bm{I}_d$ is an identity matrix. With the decoder $P(\bm{y}^t| \bm{x}^t,\bm{z}^t)$, we regard $\bm{z}^t \in \mathbb{R}^{d}$ as a low-dimensional representation of the underlying oceanographic regime at time $t$, rather than a cell-specific or cluster-specific factor. This combination of a decoder and a prior allows the framework to detect abrupt change captured by shifts in the latent prior means, which is elaborated in Section \ref{sec_learning}. While many existing change point methods are limited to detecting a specific pre-parameterized type of change, the proposed framework can capture a wide range of heterogeneous and co-occurring distributional changes through its latent representation.

\begin{figure}[!htb]
\centering

\resizebox{\columnwidth}{!}{%

\begin{tikzpicture} [on grid, state/.style={draw, minimum size=1 cm}]

\node (z1) [circle, state, dashed] {$\bm{z}^{1}$};
\node (y1) [circle, state, below = 2.2cm of z1, fill=lightgray] {$\bm{y}^{1}$};
\node (label1) [state, dashed, draw = none, above = 0.8cm of z1] {$\mathcal{N}(\bm{\mu}^1,\bm{I}_d)$};
\node (x1) [circle, state, below = 2.2cm of y1] {$\bm{x}^{1}$};

\node (z2) [circle, state, dashed, right = 4cm of z1] {$\bm{z}^{2}$};
\node (y2) [circle, state, below = 2.2cm of z2, fill=lightgray] {$\bm{y}^{2}$};
\node (label2) [state, dashed, draw = none, above = 0.8cm of z2] {$\mathcal{N}(\bm{\mu}^2,\bm{I}_d)$};
\node (x2) [circle, state, below = 2.2cm of y2] {$\bm{x}^{2}$};

\node (z3) [circle, state, dashed, right = 4cm of z2] {$\bm{z}^{3}$};
\node (y3) [circle, state, below = 2.2cm of z3, fill=lightgray] {$\bm{y}^{3}$};
\node (label3) [state, dashed, draw = none, above = 0.8cm of z3] {$\mathcal{N}(\bm{\mu}^3,\bm{I}_d)$};
\node (x3) [circle, state, below = 2.2cm of y3] {$\bm{x}^{3}$};

\node (zT) [circle, state, dashed, right = 4cm of z3] {$\bm{z}^{T}$};
\node (yT) [circle, state, below = 2.2cm of zT, fill=lightgray] {$\bm{y}^{T}$};
\node (labelT) [state, dashed, draw = none, above = 0.8cm of zT] {$\mathcal{N}(\bm{\mu}^T,\bm{I}_d)$};
\node (xT) [circle, state, below = 2.2cm of yT] {$\bm{x}^{T}$};

\path [-stealth, thick]

(z1) edge [dashed] node[right]{$P_{\bm{\phi}}(\bm{y}^1|\bm{x}^1,\bm{z}^1)$} (y1)
(x1) edge (y1)

(z2) edge [dashed] node[right]{$P_{\bm{\phi}}(\bm{y}^2|\bm{x}^2,\bm{z}^2)$} (y2)
(x2) edge (y2)

(z3) edge [dashed] node[right]{$P_{\bm{\phi}}(\bm{y}^3|\bm{x}^3,\bm{z}^3)$} (y3)
(x3) edge (y3)
 
(zT) edge [dashed] node[right]{$P_{\bm{\phi}}(\bm{y}^T|\bm{x}^T,\bm{z}^T)$} (yT)
(xT) edge (yT)

(y3) -- node[auto=false]{$\bm{\ldots}$} (yT)

(label1) edge [<->,dotted] node[above]{$\|\bm{\mu}^2-\bm{\mu}^1\|_2$} (label2)
(label2) edge [<->,dotted] node[above]{$\|\bm{\mu}^3-\bm{\mu}^2\|_2$} (label3)

;

\end{tikzpicture}
} 

\caption{An overview of prior distributions and latent space Gaussian mixture-of-experts decoder.}
\label{overview_LSGD}
\end{figure}
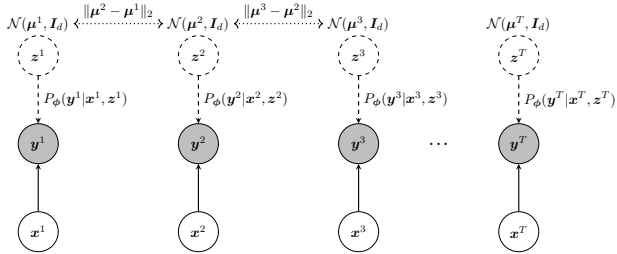

Figure \ref{overview_LSGD} gives an overview of the proposed framework. The decoder $P_{\bm{\phi}}(\bm{y}^t|\bm{z}^t, \bm{x}^t)$, with neural network parameter $\bm{\phi}$, is shared across the time points. The shaded circles in the middle denote the observed response, and the solid circles at the bottom denote the common environmental conditions as covariates shared by each cell. The dashed circles on the top denote the latent variables, where the prior means are learned to facilitate change point detection. We estimate the prior parameters $\{\bm{\mu}^t\}_{t=1}^T$ to identify change point from the optical measurements $\{\bm{y}^t\}_{t=1}^T$, and the group-fused LASSO regularization imposed on the prior parameters is described in Section \ref{sec_learning}. The simplicity of this framework is noted, without the need of encoders.

\subsection{Change Points}

In this work, the objective is to detect change points from the relationship between ocean conditions and phytoplankton species. With information abstracted to the latent representations that follow Gaussian priors, we can specify the change points through the prior parameters $\bm{\mu}^t \in \mathbb{R}^d$ for $t=1,\dots,T$. Let $\{D_r\}_{r=0}^{R+1} \subset \{1,2,\dots,T\}$ denote a set of ordered change points satisfying 
$$1 = D_0 < D_1 < \cdots < D_R < D_{R+1} = T.$$ 
These change points partition the time horizon into $R+1$ segments such that the prior parameters remain constant within each segment:
$$\bm{\mu}^{D_r} = \bm{\mu}^{D_r+1} = \cdots = \bm{\mu}^{D_{r+1}-1},\,\ r = 0,\dots,R,$$
and the prior parameters across consecutive segments are different as mean shifts:
$$\bm{\mu}^{D_r} \neq \bm{\mu}^{D_{r+1}},\,\,\ r = 0,\dots, R-1,$$
with the convention $\bm{\mu}^{D_{R+1}} = \bm{\mu}^{D_{R}}$. The multiple change point detection problem consists of recovering the set of change points $\{D_r\}_{r=1}^{R}$ from both the oceanographic covariates and optical measurements, where both the number of change points $R$ and their locations are unknown. Notationally, we denote $\bm{\mu} \in \mathbb{R}^{T \times d}$ as a matrix where the $t$-th row corresponds to $\bm{\mu}^t \in \mathbb{R}^d$.

Instead of detecting change points from replicated and clustered measurements in data space, we define change points through mean shifts in the latent space. This formulation reflects the premise that meaningful changes arise from the perturbations in underlying oceanographic conditions and their influence on species composition, as opposed to transient fluctuations in individual observations. Operating in the latent space allows the proposed method to aggregate information across replicated cells, thereby improving robustness to measurement noise, heterogeneity across clusters, and variability in mixture composition.

\section{Learning and Inference}
\label{sec_learning}

\subsection{Learning Priors from Observed Sequence}

This section develops optimization procedure to learn prior parameters for change point detection. Adapting the framework from \cite{levy2007catching,vert2010fast,lin2017sharp}, the change point detection problem for the replicated cell populations is formulated as a group-fused LASSO problem. Denote the log-likelihood of the distribution for the observed sequence as
$$l(\bm{\phi},\bm{\mu}) = \sum_{t=1}^T \sum_{b=1}^{B} \log \int P_{\bm{\phi}}(\bm{y}_b^t  | \bm{z}^t, \bm{x}^t) P_{\bm{\mu}^t}(\bm{z}^t) d \bm{z}^t.$$ 
Then we want to solve the following constrained optimization problem:
\begin{equation}
\begin{split}
\hat{\bm{\phi}},\hat{\bm{\mu}} = & \argmin_{ \bm{\phi},\bm{\mu} } -l(\bm{\phi},\bm{\mu}) + \lambda \sum_{t=1}^{T-1} \|\bm{\nu}^{t+1} - \bm{\nu}^{t}\|_2 \\ 
& \text{subject to}\ \ \bm{\mu} = \bm{\nu}.
\end{split}
\label{eqn:original_sub_to}
\end{equation}
where $\lambda > 0$ is a tuning parameter for the group-fused LASSO penalty and $\bm{\nu}^t \in \mathbb{R}^d$ denotes the $t$-th row of the slack variable $\bm{\nu} \in \mathbb{R}^{T \times d}$. Since the regularization is imposed on the prior means that relate to the likelihood, the learned priors incorporate the distributional changes into the latent space.

In particular, the group-fused LASSO penalty enforces piecewise-constant structure in the learned parameters by regularizing the multivariate total variation. The penalty promotes sparsity in the differences, while allowing multiple coordinates to shift simultaneously. When applied to the sequential differences of the prior parameters, the resulting latent representations emphasize meaningful transitions and suppress minor fluctuations.

To promote effective parameter updates, \cite{vert2010fast} introduced two variables $(\bm{\alpha}, \bm{\beta}) \in \mathbb{R}^{1 \times d} \times \mathbb{R}^{(T-1) \times d}$ to decompose the slack variable, and they are defined as
\begin{align*}
\bm{\alpha} & = \bm{\nu}^1\\
\bm{\beta}_{t,\bm{\cdot}} & = \bm{\nu}^{t+1}-\bm{\nu}^{t}
\quad
\forall\ t=1,\dots,T-1.
\end{align*}
Moreover, $\bm{\nu} \in \mathbb{R}^{T \times d}$ can be collected as $\bm{\nu} = \bm{1}_{T,1}\bm{\alpha} + \bm{D}\bm{\beta}$
where $\bm{D} \in \{0,1\}^{T \times (T-1)}$ is a design matrix with $\bm{D}_{ij} = 1$ for $i>j$ and $0$ otherwise. In essence, the decomposition of the slack variable with $(\bm{\alpha}, \bm{\beta})$ enables direct shrinkage for $\|\bm{\beta}_{t,\bm{\cdot}}\|_2$, while direct differentiation of $\|\bm{\nu}^{t+1}-\bm{\nu}^{t}\|_2 + \|\bm{\nu}^{t}-\bm{\nu}^{t-1}\|_2$ with respect to $\bm{\nu}^{t}$ would result in coupled and complicated update. In essence, the slack variable is updated in terms of $(\bm{\alpha}, \bm{\beta})$ which is easier, and then collected into $\bm{\nu} \in \mathbb{R}^{T \times d}$ for subsequent updates.

Following \cite{kei2025change}, the associated augmented Lagrangian is formulated as
\begin{equation}
\label{eqn:aug_with_alpha_beta}
\begin{split}
\mathcal{L}(\bm{\phi},\bm{\mu},\bm{\alpha}&,\bm{\beta}, \bm{u}) =  -l(\bm{\phi},\bm{\mu}) + \lambda \sum_{t=1}^{T-1}\|\bm{\beta}_{t,\bm{\cdot}}\|_2 +\\
&\frac{\rho}{2} \lVert \bm{\mu} - \bm{1}_{T,1}\bm{\alpha} - \bm{D}\bm{\beta} + \bm{u} \rVert_F^2 - \frac{\rho}{2}  \lVert \bm{u} \rVert_F^2
\end{split}
\end{equation}
where $\bm{u} \in \mathbb{R}^{T \times d}$ is the scaled dual variable and $\rho \in \mathbb{R}$ is another penalty parameter for the augmentation term. Consequently, an Alternating Direction Method of Multipliers (ADMM) procedure \citep{boyd2011distributed, zhu2017augmented, wang2019global} is used to solve the constrained optimization problem:
\begin{align}
\begin{split}
\bm{\phi}_{(a+1)},\ & \bm{\mu}_{(a+1)} = \argmin_{\bm{\phi},\bm{\mu}}\ -l(\bm{\phi},\bm{\mu}) +\\
& \frac{\rho}{2} \lVert \bm{\mu} - \bm{\nu}_{(a)} + \bm{u}_{(a)} \rVert_F^2,
\end{split}
\label{loss_phi_mu}\\
\begin{split}
\bm{\alpha}_{(a+1)},\ & \bm{\beta}_{(a+1)}  = \argmin_{\bm{\alpha},\bm{\beta}}\ \lambda \sum_{t=1}^{T-1}\|\bm{\beta}_{t,\bm{\cdot}}\|_2 +\\
& \frac{\rho}{2} \lVert \bm{\mu}_{(a+1)} - \bm{1}_{T,1}\bm{\alpha} - \bm{D}\bm{\beta} + \bm{u}_{(a)} \rVert_F^2,
\end{split}
\label{loss_nu}\\
\bm{u}_{(a+1)} & = \bm{\mu}_{(a+1)} - \bm{\nu}_{(a+1)} + \bm{u}_{(a)},
\label{loss_w}
\end{align}
where subscript $a$ denotes the ADMM iteration. The three updates are recursively implemented until certain convergence criterion is satisfied.

In particular, the solution for $\bm{\mu}^t \in \mathbb{R}^d$ at an iteration of the ADMM procedure is
\begin{equation}
\bm{\mu}^t = \frac{1}{1+\rho} \mathbb{E}_{P(\bm{z}^t|\bm{y}^t, \bm{x}^t)} ( \bm{z}^t ) + \frac{\rho}{1+\rho} (\bm{\nu}^t - \bm{u}^t)
\label{grad_mu}
\end{equation}
Moreover, the gradient with respect to the Mixture of Gaussian decoder parameter $\bm{\phi}$ is
\begin{equation}
\nabla_{\bm{\phi}}\ \mathcal{L}(\bm{\phi}, \bm{\mu}) = -\sum_{t=1}^T \mathbb{E}_{P(\bm{z}^t|\bm{y}^t, \bm{x}^t)} \Big( \nabla_{\bm{\phi}} \log P(\bm{y}^t|\bm{x}^t,\bm{z}^t) \Big).
\label{grad_decoder}
\end{equation}

The solution in (\ref{grad_mu}) and the gradient in (\ref{grad_decoder}) include the conditional expectation under the posterior density $P(\bm{z}^t|\bm{y}^t, \bm{x}^t)$. Langevin Dynamics are recruited to sample from the posterior and approximate the conditional expectations \citep{xie2017synthesizing,nijkamp2020anatomy,pang2020learning}. In particular, let subscript $k$ be the time step of the Langevin Dynamics and let $\delta$ be a small step size. The Langevin Dynamics to draw samples from the posterior is achieved by iterating the following equation:
\begin{align}
\bm{z}_{k+1}^t = \bm{z}_{k}^t + \delta \big[ \nabla_{\bm{z}^t} \log & P_{\bm{\phi}}(\bm{y}^{t}|\bm{x}^{t},\bm{z}^{t}) - \nonumber\\
& (\bm{z}_{k}^t - \bm{\mu}^t) \big] + \sqrt{2\delta} \bm{\epsilon}
\label{Langevin_sampling}
\end{align}
where $\bm{\epsilon} \sim \mathcal{N}(\bm{0},\bm{I}_d)$. Intuitively, the samples from the posterior is drawn by navigating through the posterior density in the direction of the latent variable.

Furthermore, $\bm{\beta} \in \mathbb{R}^{(T-1) \times d}$ is updated in a block coordinate descent manner \citep{bleakley2011group}, by applying the following to each row:
\begin{equation*}
\bm{\beta}_{t,\bm{\cdot}} = \frac{1}{\rho \bm{D}_{\bm{\cdot},t}^\top \bm{D}_{\bm{\cdot},t}}\Big( 1 - \frac{\lambda}{\|\bm{s}_t\|_2} \Big)_+ \bm{s}_t
\end{equation*}
where $(\cdot)_+ = \max(\cdot,0)$ and
$$\bm{s}_t = \rho \bm{D}_{\bm{\cdot},t}^\top (\bm{\mu}_{(a+1)}+\bm{u}_{(a)}-\bm{1}_{T,1}\bm{\alpha}-\bm{D}_{\bm{\cdot},-t}\bm{\beta}_{-t,\bm{\cdot}}).$$
The solution for $\bm{\alpha} \in \mathbb{R}^{1 \times d}$ is
$$\bm{\alpha} = (1/T) \bm{1}_{1,T} \cdot (\bm{\mu}_{(a+1)} + \bm{u}_{(a)} - \bm{D}\bm{\beta})^\top.$$
Once the decomposition $(\bm{\alpha}, \bm{\beta})$ are updated, the slack variable $\bm{\nu}$ is reconstructed and the scaled dual variable $\bm{u}$ is updated to complete one ADMM iteration.

\section{Change Point Localization and Model Selection}
\label{sec_guidance}

\subsection{Change Point Localization}

In this section, we describe a practical procedure for localizing change points from the estimated prior parameters. Denote the total number of ADMM iterations as $A$ and the estimated prior mean obtained at the $a$-th iteration as $\hat{\bm{\mu}}_{(a)}$. During training, we store the collection $\{\hat{\bm{\mu}}_{(a)}\}_{a=1}^A$, which tracks the evolution of the estimated prior means. At each iteration $a$, we compute the Euclidean norm of the first-order difference as
\[
\Delta \hat{\bm{\mu}}_{(a)}^t = \|\hat{\bm{\mu}}_{(a)}^{t+1} - \hat{\bm{\mu}}_{(a)}^{t}\|_2\,\,\forall\ t \in [2,T].
\]
To assess the degree of sparsity in these differences at each iteration, we calculate the empirical kurtosis:
\[
\kappa_{(a)} = 
\frac{
    \frac{1}{T-1} \sum_{t=1}^{T-1} 
    \big( \Delta \bm{\mu}_{(a)}^{t} - \text{mean}(\Delta\bm{\mu}_{(a)}) \big)^{4}
}{
    \left[
        \frac{1}{T-1} \sum_{t=1}^{T-1} 
        \big( \Delta \bm{\mu}_{(a)}^{t} - \text{mean}({\Delta\bm{\mu}}_{(a)}) \big)^{2}
    \right]^{2}
}.
\]
After training, we select the estimated prior parameter at the iteration that has maximal empirical kurtosis. Intuitively, a high kurtosis indicates that a small number of time points exhibit pronounced changes, which is consistent with the presence of abrupt shifts. To localize change points with the selected $\hat{\bm{\mu}}$, we construct a data-driven threshold defined as
\begin{equation}
\mathcal{T}_{\text{thr}} \coloneqq \text{mean}(\Delta \hat{\bm{\mu}}) + \mathcal{Z}_{\alpha} \times \text{std}(\Delta \hat{\bm{\mu}})
\label{data_driven_threshold}
\end{equation}
where $\mathcal{Z}_{\alpha}$ is the $\alpha\%$ quantile of the standard Gaussian distribution. Thus, a change point $D_r$ is identified when $\Delta \hat{\bm{\mu}}^{D_r} > \mathcal{T}_{\text{thr}}$.

\subsection{Model Selection}
\label{sec_model_selection}

The optimization problem in (\ref{eqn:original_sub_to}) involves a tuning parameter $\lambda$, which leads to different sets of detected change points when varied. We employ cross-validation to select the optimal value of $\lambda$. Specifically, the original time series data are divided into training and testing sets. The training data contains observations at odd-indexed time points, while the testing data contains those at even-indexed time points. For a fixed $\lambda$ value, the model parameters are estimated using the training data, and the log-likelihood of the Mixture of Gaussian is evaluated from the testing data. The penalty $\lambda$ with the lowest negative log-likelihood of the testing set is selected, and the model parameters are estimated again using the full dataset with the selected $\lambda$.

\section{Simulation Study with Real Covariates}
\label{simulation_study}

In this section, we evaluate the performance of the proposed method on a synthetic three-dimensional time series driven by real covariates from flow cytometry data. The clustered observations are generated from a two-component Gaussian mixture model whose parameters vary over time. In particular, at each time point $t$, we generate $N=100$ independent observations $\bm{Y}_{t,i} \in \mathbb{R}^3$.
Let
\[
\bm{X}_t = (1,\, p1_t,\, sss_t)^\top \in \mathbb{R}^3
\]
denote a vector including two real covariates, where $p1$ represents the amount of sunlight, and $sss$ is the sea surface salinity. We choose these two environmental variables because they are strongly associated with phytoplankton dynamics, and represent two fundamental ocean traits pertinent for marine microbes -- daily fluctuation in the amount of light available for photosynthesis, and the distance away from the earth with which many ocean conditions highly correlate \cite{Hyun2023}. The time horizon with $T=296$ is divided into three segments, with
$\mathcal{A}_1 = \{1,\dots,100\}$, $\mathcal{A}_2 =\{101,\dots,200\}$, and $\mathcal{A}_3 = \{201,\dots,296\}$. In this simulation, the real covariates differ by time point and are shared across all cells.

Moreover, the mixing coefficients of the Gaussian mixture model are defined as
\[
\pi_{t,k} \propto \exp(\bm{X}_t^\top \bm{\alpha}_{t,k}), \qquad k=1,2,
\]
and are normalized to unity across clusters. The vector $\bm{\alpha}_{t,k} \in \mathbb{R}^3$ is a pre-specified coefficient associated with the real covariates $\bm{X}_t$, designed to induce a change in the mixture proportions. over time. Specifically, the coefficients for Cluster 1 is $(1,0,0)^\top$ for $t \in \mathcal{A}_1 \cup \mathcal{A}_3$ and $(2,0,0)^\top$ for $t \in \mathcal{A}_2$. Those for Cluster 2 are $(0,0,0.5)^\top$ for $t \in \mathcal{A}_1 \cup \mathcal{A}_3$ and $(0,0,0)^\top$ for $t \in \mathcal{A}_2$.



Each coordinate in $\bm{Y}_{t,i}$ has a time-varying cluster-specific mean. Let $\mu_{t,d}^{(k)}$ be the mean of dimension $d \in \{1,2,3\}$ for cluster $k\in\{1,2\}$ at time~$t$. Specifically, for $d\in\{1,2\}$, the mean of Cluster~1 is
\[
\mu_{t,d}^{(k)} = p1_t,
\]
while that of Cluster~2 is
\[
\mu_{t,d}^{(k)} =
\begin{cases}
3 + 5\,p1_t, & t \in \mathcal{A}_1 \cup \mathcal{A}_3,\\[2pt]
3,          & t \in \mathcal{A}_2.
\end{cases}
\]
For $d=3$, the mean is specified as
\[
\mu_{t,3}^{(1)} = \boldsymbol{X}_t^\top(0,\,0,\,-0.5), \qquad
\mu_{t,3}^{(2)} = \boldsymbol{X}_t^\top(6,\,0,\,0.5),
\]
and does not change across segments.

Finally, each observation is generated directly from a two-component Gaussian mixture. At each time point~$t$, the distribution of
$\bm{Y}_{t,i}\in\mathbb{R}^3$ is
\[
\pi_{t,1}\, \mathcal{N}\!\bigl(\mu_{t}^{(1)},\, \sigma^2 \bm{I}_3\bigr)
\;+\;
\pi_{t,2}\, \mathcal{N}\!\bigl(\mu_{t}^{(2)},\, \sigma^2 \bm{I}_3\bigr),
\]
where $\sigma=0.5$. This yields a three-dimensional time series with covariate-driven non-stationarity with two change points located at $t=100$ and $t=200$.

Throughout, our method is denoted as CPD$_{\text{FC}}$. For the Gaussian mixture-of-experts decoder, we set the number of component to $K = 2$ and the latent dimension to $d = 3$. The decoder neural network consists of three hidden layers with $50$ neurons each; the first two layers are shared across the three branches corresponding to the mixture weights, component means, and variances. The list of tuning parameter $\lambda$ is $\{0.01,0.05,0.1,1.0\}$. For each $\lambda$, $150$ iterations of the ADMM procedure are run. The penalty parameter for the augmentation term is fixed at $\rho = 0.8$. Within each ADMM iteration, the decoder is trained for $20$ iterations using an Adam optimizer with a learning rate of $0.01$. The block coordinate descent for the slack variable is run for $20$ iterations. For Langevin dynamics, the step size is $\delta = 0.2$ and $100$ samples are drawn using $100$ iterations. After training, the change points are localized using the threshold in (\ref{data_driven_threshold}) with a significance level $\alpha = 0.99$.




We repeated the data generation process 50 times, implementing our proposed change point methodology for each and comparing with several existing methods:
\begin{enumerate}

\item \textbf{E-Divisive \citep{james2015ecp}} (\texttt{ecp}):
A nonparametric, distribution-free method that detects multiple change points using energy distances.

\item \textbf{Non-parametric PELT \citep{killick2014changepoint}} (\texttt{changepoint.np}): A penalized likelihood approach for multiple change point detection without assuming a parametric model.

\item \textbf{Breakpoint regression \citep{zeileis2002strucchange}} (\texttt{strucchange}): A structural-break procedure that estimates regression breakpoints via dynamic programming.

\item \textbf{Bayesian change point analysis \citep{erdman2008bcp}} (\texttt{bcp}): An MCMC-based method that identifies change points through posterior probabilities of mean shifts.

\end{enumerate}

Since no existing method can detect change points in replicated data, accommodations are provided to the competing methods. For each simulated sequence, we remove the covariate effects via linear regression and operate the competing methods on residuals \citep{wang2008accounting}. Specifically, let $\bm{Y}_{t,i} \in \mathbb{R}^3$ denote the three-dimensional observation at time $t$ and replicate $i$, and let $\bm{X}_t$ be the covariate vector at time $t$.
We stack the data into
\[
\bm{Y}_{\text{stack}} = \{\bm{Y}_{t,i}\} \in \mathbb{R}^{TN}, 
\qquad
\bm{X}_{\text{stack}} = \{\bm{X}_t\} \in \mathbb{R}^{TN \times 3},
\]
and fit a linear regression as
\[
\bm{Y}_{\text{stack}}^{(j)} = \bm{X}_{\text{stack}}\bm{\beta}^{(j)} 
+ \bm{\varepsilon}^{(j)}.
\]
The fitted residuals are then reshaped to their original time indices, yielding a vector of residual 
$\bm{r} = \{r_{t,i}^{(j)}\} \in \mathbb{R}^{TN}$ for \(j=1,2,3\).

For methods requiring univariate time series, we aggregate the residuals at
each time point via the squared norm
\[
r_t = 
\sum_{i=1}^N \sum_{j=1}^3 
\bigl( r_{t,i}^{(j)} \bigr)^2,
\]
producing a single time series $\{r_t\}_{t=1}^T$ with length $T$ that summarizes the total
residual variation after regressing out the covariate effects.

The four competing methods are applied to these derived features as follows. The \texttt{strucchange} and \texttt{bcp} methods use univariate series $\{r_t\}_{t=1}^T \subset \mathbb{R}$, detecting mean shifts in the residuals. The PELT method from \texttt{changepoint.np} is applied to $\{r_t\}_{t=1}^T$ using a non-parametric cost function. For the multivariate method \texttt{ecp}, we use the three-dimensional average residual vector at each time,
\[
\mathbf{m}_t
= \frac{1}{N}
\sum_{i=1}^N 
\bigl(
\mathrm{resid}_{t,i}^{(1)},\,
\mathrm{resid}_{t,i}^{(2)},\,
\mathrm{resid}_{t,i}^{(3)}
\bigr),
\]
and applied E-divisive to the sequence $\{\mathbf{m}_t\}_{t=1}^T$.

To evaluate the performance of the change point detection methods, we use the following metrics on the detected set of locations $\widehat{\mathbf{c}} = \{\widehat{c}_1,\dots,
\widehat{c}_K\}$ with the true set $\mathbf{c}^\star=\{c_1^\star,\dots, c_{K^\star}^\star\}$. 



\noindent\textbf{(1) False positives (FP).}
Estimated change points that do not match any true change point within the
tolerance~$\tau$:
\[
\mathrm{FP}
= \sum_{k}\mathbf{1}\!\left(
  \min_{j}|\widehat{c}_k - c_j^\star| > \tau
\right).
\]

\noindent\textbf{(2) False negatives (FN).}
True change points that are not detected within~$\tau$:
\[
\mathrm{FN}
= \sum_{j}\mathbf{1}\!\left(
  \min_{k}|\widehat{c}_k - c_j^\star| > \tau
\right).
\]
\noindent\textbf{(3) $D_{\mathrm{t\to e}}$.}
Worst-case distance from each true change point to its nearest estimate:
\[
D_{\mathrm{t\to e}}
= \max_{j}\;\min_{k}|\widehat{c}_k - c_j^\star|.
\]
\noindent\textbf{(4) $D_{\mathrm{e\to t}}$.}
Worst-case distance from each estimated change point to its nearest true one:
\[
D_{\mathrm{e\to t}}
= \max_{k}\;\min_{j}|\widehat{c}_k - c_j^\star|.
\]
\noindent\textbf{(5) Count Error (CE).}
Difference in the number of detected and true change points:
\[
\mathrm{CE} = \bigl|N_{cp} - N^\star_{cp}\bigr|,
\]
where $N_{cp}$ and $N^\star_{cp}$ denotes the true and estimated number of change points, respectively.

\noindent\textbf{(6) Cover Score (CS).}
A segmentation-level score based on the best Jaccard overlap between true
and estimated segments:
$$C(\mathcal{G},\mathcal{G'}) = \frac{1}{T} \sum_{\mathcal{A} \in \mathcal{G}} |\mathcal{A}| \cdot \max_{\mathcal{A'} \in \mathcal{G'}} \frac{|\mathcal{A} \cap \mathcal{A'}|}{|\mathcal{A} \cup \mathcal{A'}|}$$
where $\mathcal{G}$ and $\mathcal{G}'$ are the respective ground truth and estimated segmentation.

Table \ref{tab:sim} reports the averages over 50 simulated datasets for the six evaluation metrics. Overall, our method achieves lower false positive (FP) and false negative (FN) rates, as well as a lower worst-case distance $D_{\mathrm{t\to e}}$ than all competing approaches. In addition, our approach attains a higher cover score (CS), indicating accurate recovery of the true change point locations on average. For the worst case distance $D_{\mathrm{e\to t}}$ and the change point count error (CE), our method ranks second among all methods, suggesting that it  detects slightly more change points than the ground truth. In contrast, breakpoint regression and E-divisive methods tend to underestimate the number of change points, reflecting a conservative segmentation strategy that reduces the risk of spurious detections but leads to higher false negative rate. Taken together, these results demonstrate that the proposed method performs well across the evaluation metrics.

\begin{table}[!ht]
\caption{Comparison of change point detection performance across methods. The best metric is in bold.}
\label{tab:sim}
\centering
\resizebox{\columnwidth}{!}{
\begin{sc}
\begin{tabular}{cccccccc}
\toprule
Method & FP$\downarrow$ & FN$\downarrow$ & $D_{\mathrm{t\to e}}\downarrow$ & $D_{\mathrm{e\to t}}\downarrow$ & CE$\downarrow$ & CS$\uparrow$ \\
\midrule
CPD$_{\text{FC}}$  & \textbf{0.60} & \textbf{0.04} & \textbf{2.98} & 39.72 & 0.56 & \textbf{0.94} \\
bcp             & 3.92 & 1.00 & 95.88 & 96.64 & 2.92 & 0.46 \\
strucchange     & 1.00 & 1.00 & 47.12 & 47.12 & \textbf{0.00} & 0.70 \\
ecp             & 1.20 & 0.92 & 12.10 & \textbf{18.32} & 0.28 & 0.89 \\
changepoint.np  & 1.04 & 0.04 & 8.80 & 58.52 & 1.00 & 0.80 \\
\bottomrule
\end{tabular}
\end{sc}
}
\end{table}



\section{Application to Flow Cytometry Data}

\begin{figure}[h!]
\centering
\includegraphics[width = 1\linewidth, clip, trim={1cm 5cm .8cm 4.5cm}]{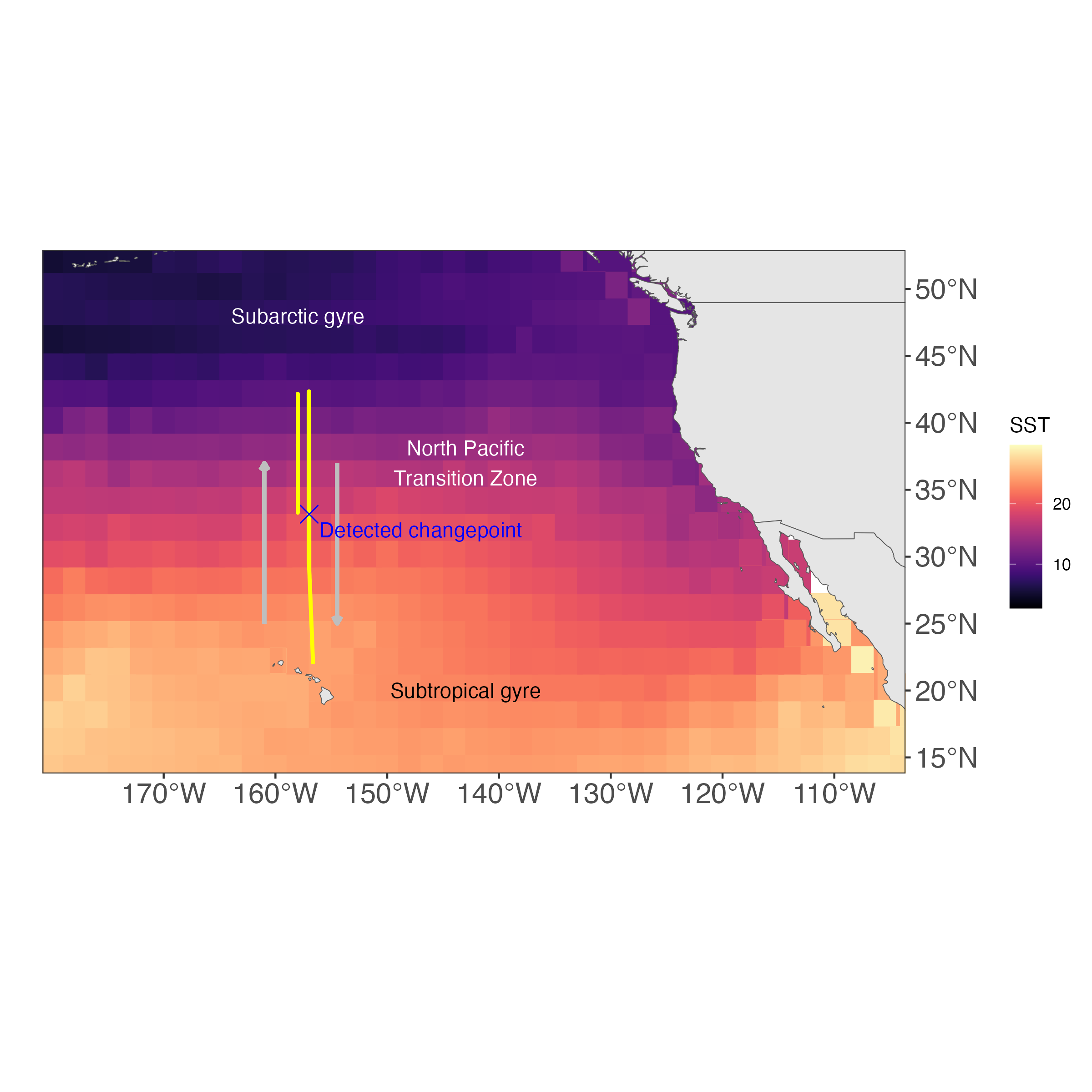}
\caption{Map of the research cruise trajectory in the North Pacific Ocean shown in yellow (the upward trajectory and downward trajectory are slightly separated for visibility). The research cruise (Gradients 2) traversed two different bodies of ocean water ---``subtropical gyre'' in the south and ``subarctic gyre'' up north. Our proposed method applied to this dataset detected a change point at about 33.2 degrees North, in latitude, which is roughly consistent with several other change points found in the literature.}
\label{fig:map}
\end{figure}

In this section, we apply the proposed method to a real flow cytometry data.  Figure~\ref{fig:map} shows a map of the research cruise trajectory where data was collected in the North Pacific Ocean --- in yellow, the upward trajectory and downward trajectory are plotted with a small artificial separation. The research cruise (Gradients 2) traveled up then down from Hawaii over the course of about two weeks, and traversed two different bodies of ocean water --- the southern waters called the ``subtropical gyre'' and northern waters called the ``subarctic gyre''. The Seaflow flow cytometer \citep{Swalwell2011} was used to collect particle-level data for our experiment. In this plot, we mark in blue the detected change point using our proposed method --- at about 33.2 degrees North in latitude. We  further discuss this exciting result at the end of this section.

\label{Application}
\begin{figure*}[htbp]
    \centering
    \includegraphics[width=0.9\linewidth]{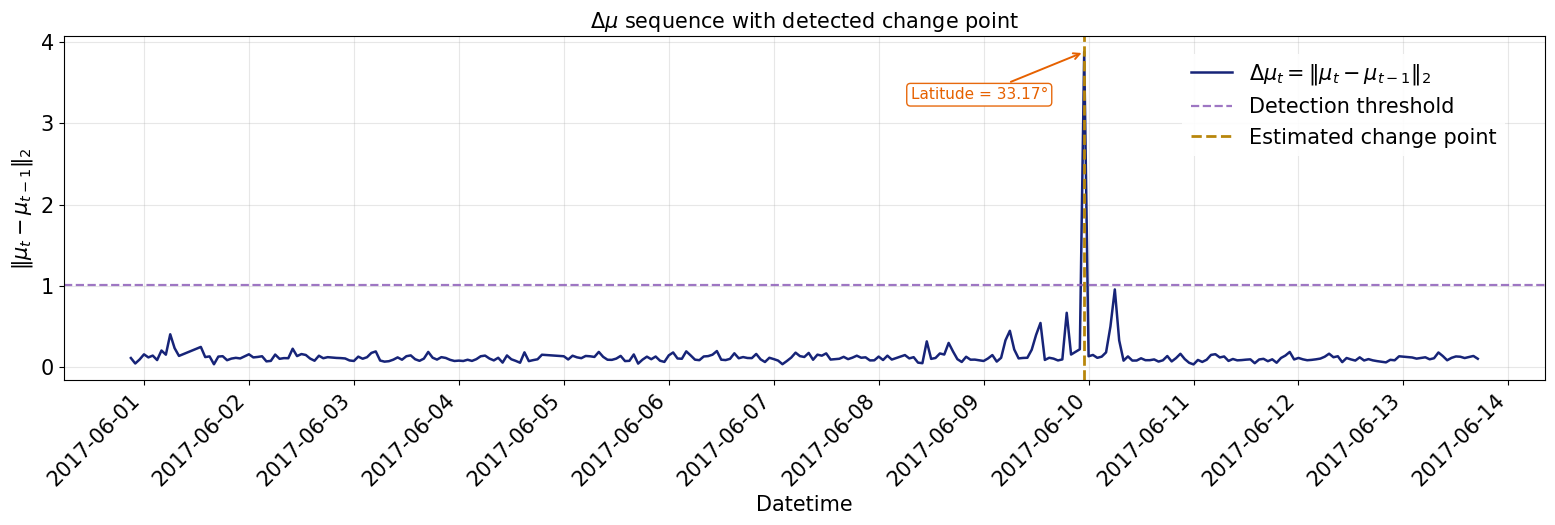}
    \caption{An illustration of the estimated $\Delta\mu$ for the flow cytometry data. The detected change points are indicated by the red vertical line. The detection threshold is displayed by the horizontal line.}
    \label{fig:application}
\end{figure*}


Our dataset of interest consists of $296$ hourly time points spanning from June 1, 2017 to June 14, 2017. At each time point, the number of measured phytoplankton cells varies. A total of 39 covariates are available,  and cells measured at the same time point share a common covariate vector. In addition, a biomass measurement is observed for each data point, which serves as a weight reflecting its relative contribution.

In the original dataset, there are $296$ times points -- each containing one full hours' worth of repeated microbial observations (cytometric particles). The number of observations at each time varies from about $1,986$ to $20,181$. Due to the large size of the dataset, we perform weighted resampling at each time point based on the biomass values. Specifically, $B = 500$ observations are sampled at each time point. This sample size was chosen to balance cell diversity and computational efficiency.


For particulars, we set the number of mixture components to $K = 15$ and the latent dimension to $d = 5$. The decoder neural networks consist of three hidden layers with $50$ neurons each. The first two layers are shared across the three branches that correspond to the component weights, means, and variances. The ADMM penalty parameter $\lambda$ is selected from $\{0.01, 0.05, 0.1, 1.0\}$ via cross validation. For each $\lambda$, $200$ iterations of the ADMM procedure are run. The decoder is trained for $10$ iterations with Adam optimizer using a learning rate of $0.001$. The penalty parameter for the augmentation term is fixed at $\rho = 0.8$, and the block coordinate descent is run for $10$ iterations. For Langevin dynamics sampling, the step size is $0.2$ with $200$ iterations and $500$ samples. Change point localization is based on the first-order differences of the estimated mean, with a significance level of $0.99$ for thresholding. Figure~\ref{fig:application} displays the estimated $\Delta\bm{\mu}$ over time.


Our method detected a changepoint at 33.2°N, which is quite comparable to various biological and physical change points that are found in the literature studying this North Pacific transect region. In particular, it is within 1 degrees of two {\it biological} changepoints (approximately 33.1°N and 33.7°N) that were also detected by \cite{Jones2021} (Figure 6, left panel), which analyzed flow cytometry data from a similar research cruise that went out April 20 through May 2, 2016, about a year and a month earlier. Also, the famed Longhurst framework \cite{Longhurst_1995, longhurst-book}, which classically studies marine provinces, defines a division between the two North Pacific Provinces (NPPF (North Pacific Polar Frontal region/zone) and NPTG (North Pacific Tropical Gyre) at about 34.4°N latitude along the vertical trajectory of this cruise.
Additionally, \cite{Follett2022} studies the collapse of a certain prominent photosynthetic microbe species (called {\it Prochlorococcus}), which occurs at about 30.4°N, a few degrees south of where we detect the point.
Lastly, although not explicitly detecting change points, two studies
\cite{hyun2023modeling, hyun2025trendfilteredmixtureexperts}
each also found estimates of (one gradual, one more sudden) rise in {\it Prochlorococcus} abundance around the same latitude at the beginning of June 8th, remarkably similar to our detected change point.

\section{Discussion}
\label{sec_discussion}

This paper proposes a latent space Gaussian mixture-of-experts model for change point detection in replicated phytoplankton observations with environmental covariates. To the best of our knowledge, existing change point detection methods are not designed to handle replicated and clustered single-cell measurements, whose distributions evolve along with the covariates. By mapping cytometric responses into low-dimensional representations, the proposed method identifies change points through mean shifts in the latent space, thereby capturing changes in the underlying ocean conditions. Simulation studies demonstrate that the proposed method performs well and real data analysis produces results consistent with existing scientific understanding.

Several extensions are worth exploring. Allowing the number of mixture components to vary over time may improve flexibility, since new cell populations may emerge and existing populations may diminish as a cruise traverse through different regions. It would also be of interest to extend the model to detect gradual, not abrupt, changes in the underlying mixture distribution, by adopting alternative penalization strategies. Finally, incorporating more structured or parametric change point models, such as linear or constrained expert components, may improve interpretability and offer a promising direction for scientific discovery.



\section*{Impact Statement}

This paper presents work whose goal is to advance the field of change point detection in replicated and clustered cell populations with covariates. There are many potential environmental consequences of our work, including development of new capabilities for mapping ocean provinces and understanding phytoplankton population dynamics.

\newpage
\bibliography{reference}
\bibliographystyle{icml2026}


\end{document}